\documentclass{emulateapj}
\usepackage[utf8]{inputenc}
\usepackage[T1]{fontenc}
\usepackage{graphicx}
\usepackage{longtable}
\usepackage{soul}
\usepackage{hyperref}
\usepackage{amsmath, amssymb}
\usepackage{ulem}
\usepackage{natbib}

\shortauthors{Hull et al.}
\shorttitle{Primary Beam Shape Calibration}

\begin{document}

\newcommand{\etal}{{\it et al.\ }}

\slugcomment{\textit{Accepted by P.A.S.P.}}

\title{Primary Beam Shape Calibration from Mosaicked, Interferometric Observations}

\author{Charles L. H. Hull\altaffilmark{1}, Geoffrey C. Bower\altaffilmark{1}, Steve Croft\altaffilmark{1}, Peter K. G. Williams\altaffilmark{1}, Casey Law\altaffilmark{1}, David Whysong\altaffilmark{1}}
\email{chat@astro.berkeley.edu}

\altaffiltext{1}{Astronomy Department \& Radio Astronomy Laboratory, University of California, Berkeley, CA 94720}

\begin{abstract}
Image quality in mosaicked observations from interferometric radio telescopes is strongly dependent on the accuracy with which the antenna primary beam is calibrated. The next generation of radio telescope arrays such as the Allen Telescope Array (ATA) and the Square Kilometer Array (SKA) have key science goals that involve making large mosaicked observations filled with bright point sources. We present a new method for calibrating the shape of the telescope's mean primary beam that uses the multiple redundant observations of these bright sources in the mosaic. The method has an analytical solution for simple Gaussian beam shapes but can also be applied to more complex beam shapes through $\chi^2$ minimization. One major benefit of this simple, conceptually clean method is that it makes use of the science data for calibration purposes, thus saving telescope time and improving accuracy through simultaneous calibration and observation. We apply the method both to 1.43 GHz data taken during the ATA Twenty Centimeter Survey (ATATS) and to 3.14 GHz data taken during the ATA's Pi Gigahertz Sky Survey (PiGSS). We find that the beam's calculated full width at half maximum (FWHM) values are consistent with the theoretical values, the values measured by several independent methods, and the values from the simulation we use to demonstrate the effectiveness of our method on data from future telescopes such as the expanded ATA and the SKA.  These results are preliminary, and can be expanded upon by fitting more complex beam shapes.  We also investigate, by way of a simulation, the dependence of the accuracy of the telescope's FWHM on antenna number.  We find that the uncertainty returned by our fitting method is inversely proportional to the number of antennas in the array.
\end{abstract}


\section{Introduction}
\label{sec:intro}

The ATA is a "Large Number of Small Dishes" (LNSD) interferometric array located in Northern California at the Hat Creek Radio Observatory (HCRO) of the University of California, Berkeley.  The telescope is designed to be highly effective for commensal surveys of conventional radio astronomy projects and SETI (Search for Extra-Terrestrial Intelligence) targets at centimeter wavelengths, and currently consists of 42 6-meter dishes with continuous frequency coverage over 0.5--10 GHz \citep{2009IEEEP..97.1438W, 2010ApJ...710.1462W}.  The ATA has a wide field of view (4.7 deg$^2$ at 1.43 GHz), which allows for rapid surveying of large areas of the sky.  Since the point-source sensitivity of wide-field surveys requiring multiple pointings is proportional to $ND$ (as opposed to the total collecting area $ND^2$), where $N$ is the number of dishes and $D$ is the dish diameter, the LNSD design was the natural choice for the ATA \citep{1993ARA&A..31..297S}.

Since the ATA is a wide-field survey instrument, the construction of mosaics---large images created by stitching together snapshots from individual pointings---is of prime importance.  In order to account for the attenuation of sources with distance from the center of the primary beam (i.e. the pointing center), and to correct for this effect when mosaicking the images, we must know the shape of the telescope's primary beam.

The results of not properly characterizing the beam can be devastating.  \citet{1993A&A...271..697C} discuss the effects of poor beam characterization on image fidelity, noting that deviations from the primary beam model of a telescope should be small down to or below the $\sim$7\% level.  \citet{SKA_Memo_103} show that if the actual primary beam pattern deviates from the canonical pattern by just a few percent, the image fidelity can be reduced by more than two orders of magnitude.  

Our goal is to self consistently measure the FWHM with science data, which allows us to monitor the beam shape in real time and at the current observing frequency.  While work such as the holography mapping study described in \citet{Harp:ATA_beam_holography} is invaluable as a thorough initial calibration of the beam shape of the telescope, it is time-consuming, and cannot be repeated regularly.  Additionally, holography mapping is performed by training the telescope on strong satellite emission, which may be at a frequency different from that being used for observations.  The real-time calibration work we present here is complementary to such detailed studies.

\citet{Welch:ATA_memo_66} predicted the gain of the ATA antennas to be

\begin{equation}
G(\theta) \propto \left[ \frac{J_{\textrm{1}}\left[ \left( 6\pi / \lambda \right) \sin \theta \right]}{\left( 6\pi / \lambda \right) \sin \theta} + 25.40 \, \frac{J_{\textrm{2.9}}\left[ \left( 6\pi / \lambda \right) \sin \theta \right]}{\left( \left( 6\pi / \lambda \right) \sin \theta \right)^{2.9}} \right]^2 \\ \,\, ,
\label{eqn:bessel_beam}
\end{equation}

\noindent
where $\theta$ is the angular distance from the pointing center.  Furthermore, the beam can be approximated by a symmetric Gaussian of the form

\begin{equation}
G(\theta) \propto \exp{\left[-4 \ln 2 \, \left(\frac{\theta}{\Theta_{1/2}}\right)^2\right]} \,\, ,
\label{eqn:gauss_beam}
\end{equation}

\noindent
where $\Theta_{1/2}$ is the theoretical full width at half maximum (FWHM) of the Gaussian beam.  

\begin{figure*}[hbt!]
\begin{center}
\plottwo{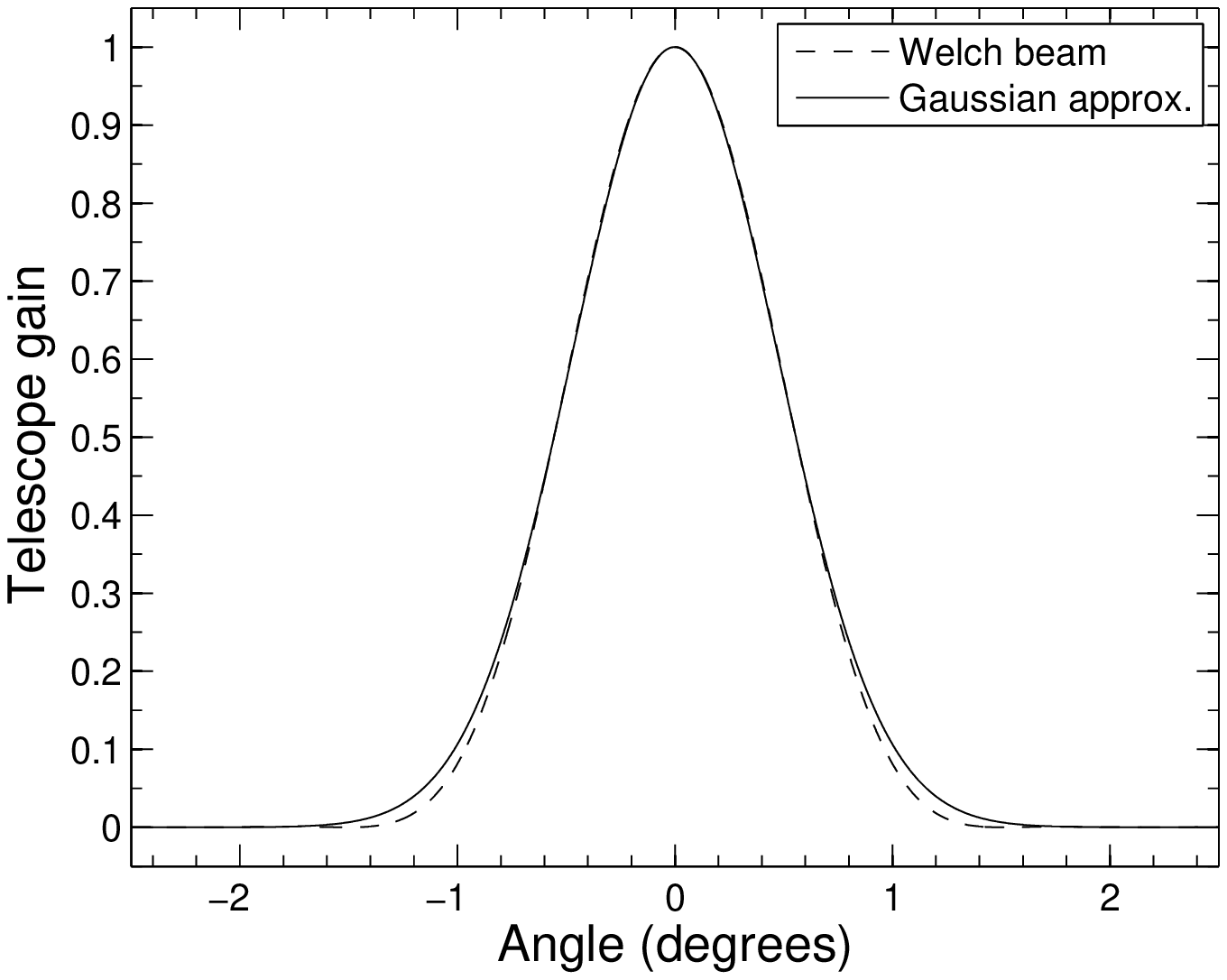}{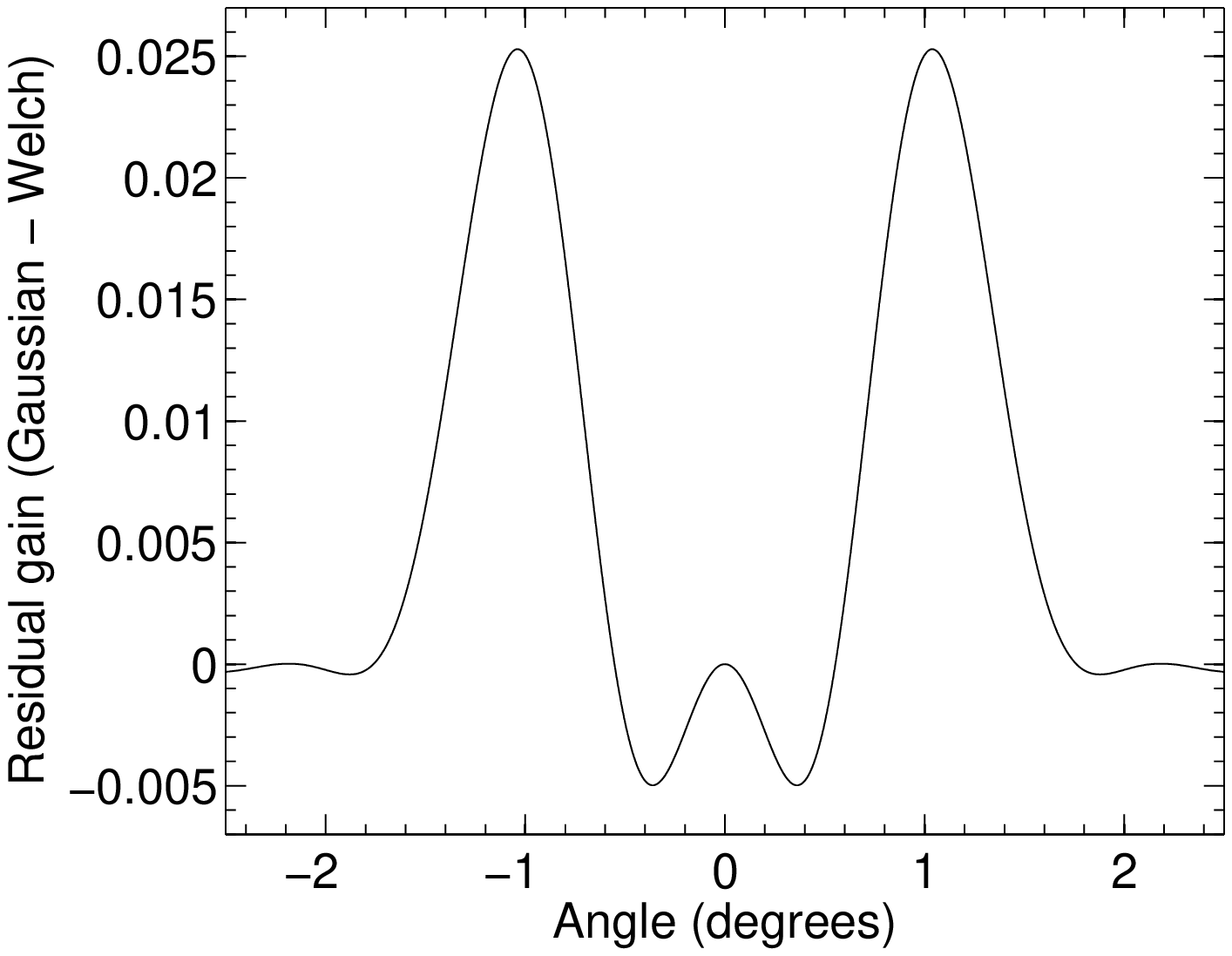}
\caption{The beam shape from \citet{Welch:ATA_memo_66} and the Gaussian beam approximation (left), and the residual gain after taking the difference between the Gaussian approximation and the Welch beam (right).  The plotted beams correspond to the PiGSS frequency of 3.14 GHz.}
\label{fig:welch_vs_gauss}
\end{center}
\end{figure*}

Both the beam from Eqn. \ref{eqn:bessel_beam} and the Gaussian beam approximation from Eqn. \ref{eqn:gauss_beam} are plotted in Figure \ref{fig:welch_vs_gauss}, along with the residual between them.  The residual peaks at 2.5\% at a gain of less than 10\%; the beam shapes differ by less than 0.5\% inside the beam's half-power points.  These small differences suggest that the Gaussian approximation from Eqn. \ref{eqn:gauss_beam} is an acceptable alternative to the more complicated beam shape from Eqn. \ref{eqn:bessel_beam}.

Scaling the predicted $\Theta_{1/2}$ with frequency in units of GHz gives

\begin{equation}
\Theta_{1/2} = \frac{\Theta_0}{f(\textrm{GHz})} \,\, ,
\label{eqn:FWHM}
\end{equation}

\noindent
where the canonical value of $\Theta_0$ is 3.50$^\circ$.

Equation \ref{eqn:FWHM} gives us an expected FWHM of the mean primary beam of the ATA (the average of the primary beams of each dish) of $2.45^{\circ}$ at the ATATS frequency of 1.43 GHz \citep{2010ApJ...719...45C}, and $1.11^{\circ}$ at the PiGSS frequency of 3.14 GHz \citep{2010arXiv1009.4443B}.  The actual shape of the beam may not match this prediction, however.  In order to mosaic the large numbers of snapshots successfully, we must properly characterize the beam.  

In $\S$\ref{sec:method}, we explain our two methods for characterizing the FWHM of the telescope's mean primary beam, after which, in $\S$\ref{sec:data}, we describe the data to which we applied our methods.  We then report our results in $\S$\ref{sec:results}, and we conclude in $\S$\ref{sec:ska} with a discussion of a simulation that applies our methods to simulated data of future telescopes such as the expanded, 350-dish ATA and the 3000-dish SKA (SKA-3000).

\section{Description of Methods}
\label{sec:method}

\begin{figure}
\begin{center}
\plotone{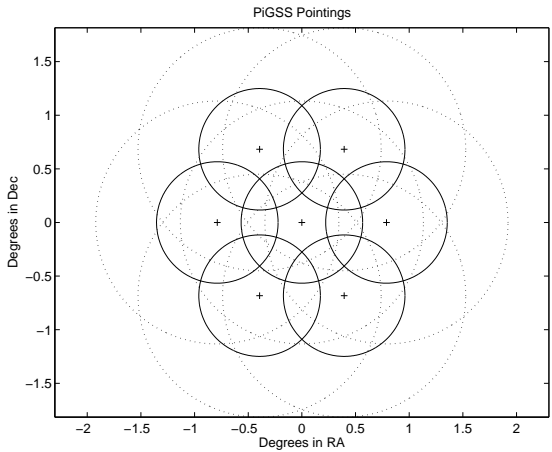}
\caption{An illustration of seven overlapping PiGSS pointings.  The crosses are the pointing centers, the solid circles denote the primary beam's half-power radius, and the dotted circles denote twice the half-power radius.  This hexagonal close-packed arrangement of pointings is typical of other all-sky surveys, including ATATS.}
\label{fig:pointings}
\end{center}
\end{figure}

As pictured in Fig. \ref{fig:pointings}, pointings used in an ATA mosaic overlap significantly, having been arranged to approximate a hexagonal close-packed grid.  The distances between pointing centers are determined by Nyquist sampling, and the convergence toward the poles of lines of constant RA has been accounted for.  This is a typical pointing grid used in many all-sky surveys with the ATA and other telescopes.

Sources can appear in different locations within the beam in several adjacent, overlapping pointings.  Bright sources ($\sim$1 Jy) are sometimes detectable in as many as six or seven adjacent pointings, depending on where in the primary beam they fall in each pointing.  Dimmer sources tend only to be detected in two or three pointings.  We use the multiple appearances of each source to constrain the beam's FWHM by comparing the flux densities of the source in adjacent pointings.

To find matching pairs, we compare each detected source with every other source, declaring a match when the difference between the two sources' reported positions is less than the typical pointing error of one arc minute.  We accumulate 4,304 matched pairs in the ATATS data, and 824 pairs in the PiGSS data.  After matching sources, we attempt to constrain the beam's FWHM using two methods of parameter fitting.

This method assumes that the sky is stationary on the time scale of observations.  Transient and variable sources, therefore, will reduce the accuracy of the overall result.  Transient sources --- defined as those that appear in a single pointing --- can be rejected on the absence of a source match in another and will then have a minimal effect as long as source confusion is not important.  Variable sources present a more complex problem.  It is likely that strongly variable sources can be rejected with a technique such as median filtering.  But sources with moderate variability on the timescale of the observations will be difficult to distinguish from distortions in the primary beam shape.  Fortunately, the number of variable sources is a relatively small fraction (less than 1\% at mJy sensitivity) of the total number of radio sources \citep{2007ApJ...666..346B,2010ApJ...719...45C,2010arXiv1009.4443B}.

We must also emphasize that the true coordinates for the primary beam are (Az, El), while these measurements give results for celestial coordinates (RA, DEC).  A more sophisticated model would require translating (RA, DEC) to (Az, El), especially if we wanted to work with long tracks of data.  Our current method is appropriate for situations where the beam is circularly symmetric, and when working with the snapshot observations typical of high-cadence surveys such as ATATS and PiGSS.

\subsection{Method 1: two-point Gaussian fitting}
\label{sec:two_point}

Assuming only a radial sensitivity dependence within the telescope's primary beam, we can calculate the FWHM of a Gaussian that we fit to each pair of matched sources.  We represent the measured flux densities in the following way:
	
\begin{align} 
S_i &= S_{\textrm{true}} \, G(\theta_i) \label{eqn:S_i} \,\, ,
\end{align}

\noindent
where $S_{\textrm{true}}$ is the absolute flux density of the source (unknown to us); $S_i$ is the observed flux density of a source in a given pointing; $\theta_i$ is the distance of the source from its respective pointing center; and $G(\theta)$ is the gain of the telescope as a function of the distance from the pointing center.  

If we assume the gain is a circular Gaussian with a given standard deviation $\sigma$, then the observed flux densities are

\begin{align} 
S_i &= S_{\textrm{true}} \, \exp{\left(- \frac{\theta_i^2}{2 \sigma^2}\right)} \label{eqn:S_i_gauss} \,\, .
\end{align}

Solving for $\Theta_{1/2}$ for a given pair of sources gives

\begin{equation}
\Theta_{1/2} = \sqrt \frac{4 \ln{2} \, (\theta_2^2 - \theta_1^2)}{\log \, (S_1/S_2)} \label{eqn:r_FWHM} \,\, .
\end{equation}

Our analytical method for calculating FWHM values using Equation \ref{eqn:r_FWHM} breaks down under two circumstances: when $S_1/S_2 \approx 1$ (which causes us to divide by zero), and when $\theta_1/\theta_2 \approx 1$ (which results in a FWHM of zero).  (These two cases are essentially the same, since when $\theta_1 \approx \theta_2$, it follows that $S_1 \approx S_2$.  This is assuming a symmetric primary beam, and assuming that the source has a constant flux density and hasn't been mismatched.)  This is unfortunate, because the very places where Equation \ref{eqn:r_FWHM} breaks down are where we would expect the largest number of pairs to be detected: that is, where the source appears roughly the same distance from each pointing center and is not attenuated too much in either pointing. 

The advantages of this method are that it calculates a FWHM measurement for each pair of matches; for a purely Gaussian beam shape, the analytical solution for the FWHM is simple to calculate; and the method's accuracy should increase as the number of observed sources is increased.  The disadvantages, however, also become apparent in our discussion above.

\subsection{Method 2: $\chi^2$ minimization}
\label{sec:chi2}

Instead of solving for a separate FWHM for each flux density pair, this approach allows us to correct all of the data at once with one best-fit FWHM.  We correct the measured flux densities of each pair of matched sources to their (presumed) true values by multiplying by the inverse of the gain, which is defined according to Equation \ref{eqn:gauss_beam}, and depends on the primary beam's FWHM.  To arrive at the best-fit FWHM value, we assume a FWHM for the beam, correct all of the flux density pairs using that FWHM, and calculate the reduced-$\chi^2$ ($\chi^2_\nu$) using the following formula:  

\begin{equation}
\chi^2_\nu = \frac{1}{\nu} \sum \limits_{i = 0}^N \frac{ \left( S_{1,\textrm{true},i} - S_{2,\textrm{true},i} \right)^2 }{ \Delta S_{1,i}^2 + \Delta S_{2,i}^2 } 
\end{equation}

\noindent
where $\nu$ is the number of degrees of freedom, $N$ is the number of matched pairs, $S_{1,\textrm{true},i}$ and $S_{2,\textrm{true},i}$ are the corrected flux densities, and $\Delta S_{1,i}^2$ and $\Delta S_{2,i}^2$ are the uncertainties in flux density.

An important advantage of this method is that it, unlike the two-point method, it uses all of the data to determine the FWHM of the primary beam.  Additionally, we can use this method to fit any arbitrarily complex beam shape, not just simple shapes like the purely Gaussian shape $G(\theta)$ that we have assumed above.  For example, the method could easily be extended to fit other parameters such as the beam's ellipticity and angle on the sky.

\section{Description of Data}
\label{sec:data}

\subsection{ATATS}
\label{sec:data:ATATS}

We use both methods described above to determine a value for the FWHM for the primary beam of the ATA, using data from the ATA Twenty-centimeter survey \citep[ATATS; ][]{2010ApJ...719...45C}.  ATATS is a multi-epoch, 690 deg$^2$ survey at 1.43 GHz, and is designed to search for transients as well as to verify the capabilities of the ATA to form large mosaics.  ATATS is centered on the $\sim$10 deg$^2$ Bo\"{o}tes deep field \citep{1999ASPC..191..111J}.  The data were reduced using a software suite developed for reduction of ATA data, RAPID \citep[Rapid Automated Processing and Imaging of Data;][]{2009AAS...21460106K}.  RAPID calls MIRIAD \citep{1995adass...4..433S} tasks using C shell scripts, and performs RFI excision, visibility calibration, phase self-calibration, and amplitude flagging.  

The images were deconvolved using the MIRIAD task "clean". The number of iterations was set using an "intelligent clean" algorithm implemented by RAPID. The flux densities of point sources in the field, and the measured RMS noise, are used to calculate the appropriate number of clean components to clean down to the $3 \sigma$ level. At this stage, RAPID checks for the presence of point sources in the residual image, and if any are found, the number of clean iterations is iteratively increased until the residual images are consistent with noise.

The sources we used in our analysis were 5623 sources from a catalog created by running the MIRIAD task SFIND on maps of individual pointings from ATATS Epoch 12, taken on 2009 April 3 \citep{2010ApJ...719...45C}.  Each of the 253 pointings is a snapshot with one minute of integration time.  The overlapping of the individual pointings, as described in $\S$\ref{sec:method} and depicted in Fig. \ref{fig:pointings}, allowed us to detect the same source in several different snapshots.  These data were taken as part of a pilot study; ongoing improvements to the array and to the data analysis routines are resulting in continuing enhancements in image quality and data consistency.

\subsection{PiGSS}
\label{sec:data:PiGSS}

The Pi Gigahertz Sky Survey \citep[PiGSS;][]{2010arXiv1009.4443B} is a radio continuum survey at 3.14 GHz, which observes the North Galactic Cap and other selected 10 deg$^2$ fields.   These data were processed using methods similar to those used on the ATATS data; the images are restored with a circular synthesized beam of 100" FWHM.  The data used in our analysis come from the individual images comprising a seven-pointing mosaic of the Bo\"{o}tes deep field.  These images were obtained from 75 daily observations, each of which included between two and six snapshots of each pointing, providing very good coverage of the $(u,v)$ plane.  Maps of each pointing were produced with a size of $4.26^{\circ}$ on a side, much larger than the primary beam FWHM of $\sim$$1.1^{\circ}$.  A total of 672 sources were identified in the individual pointings using the MIRIAD task SFIND.

\section{Results}
\label{sec:results}

\subsection{Two-point method}

\begin{figure*}[hbt!]
\begin{center}
\plottwo{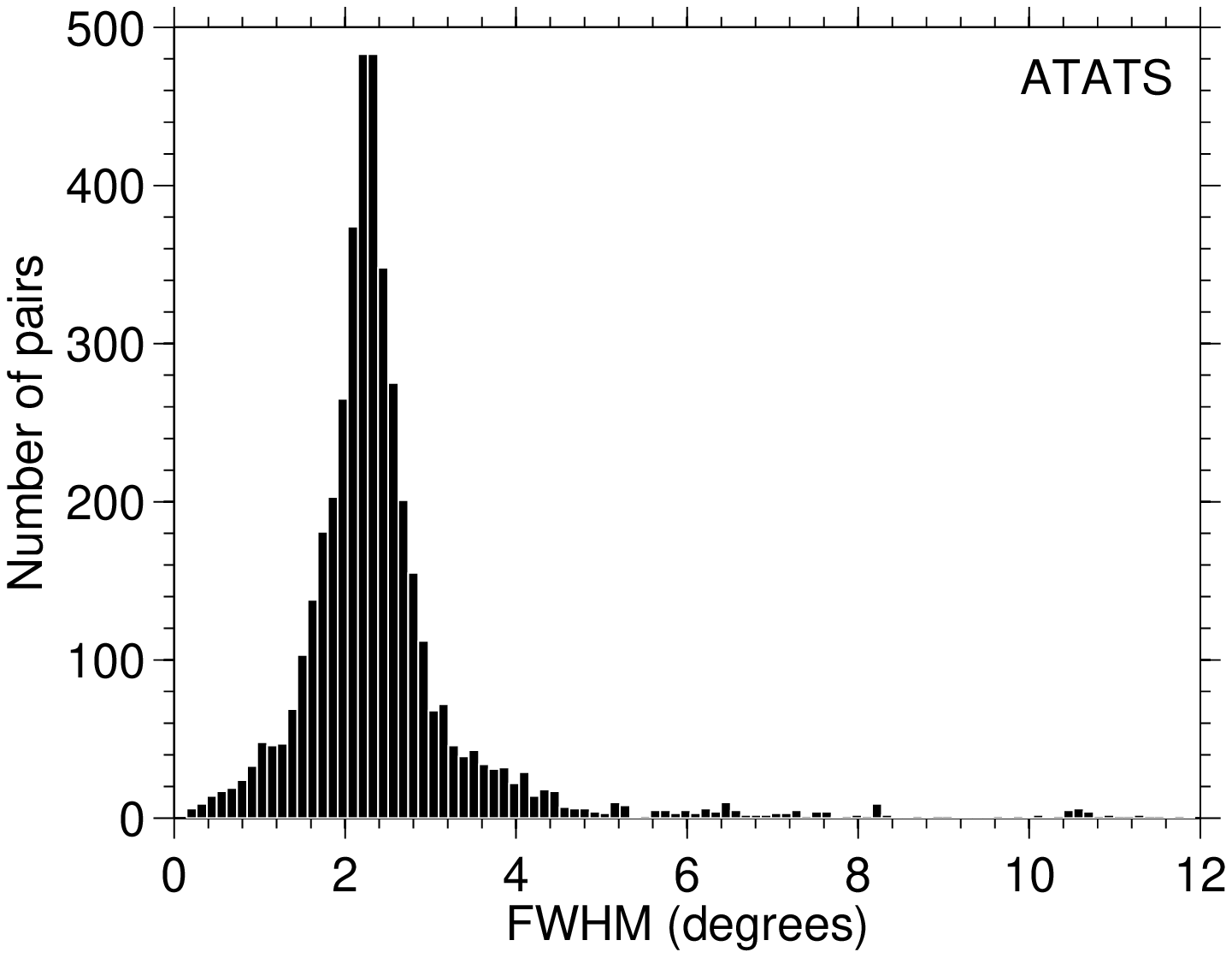}{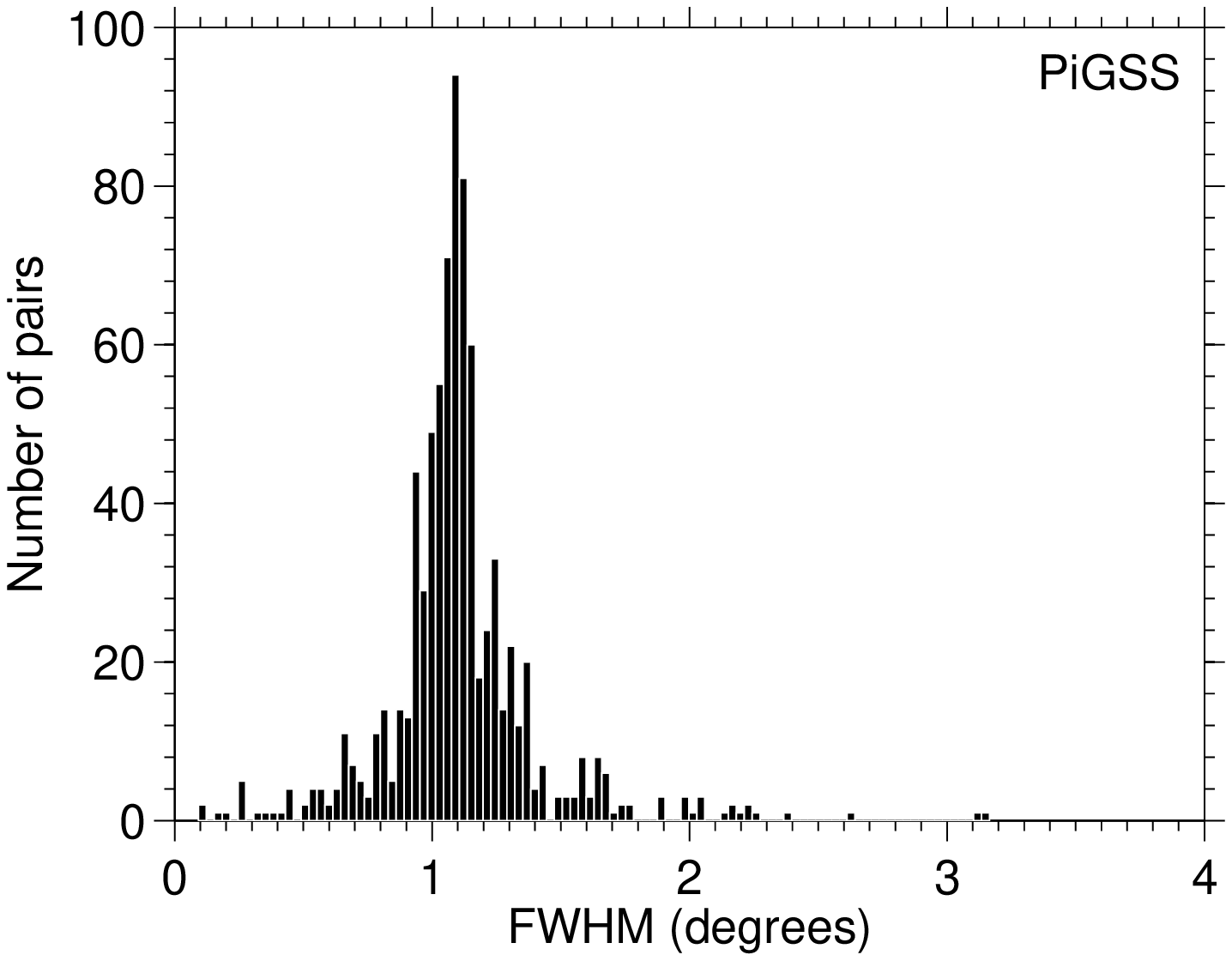}
\caption{A histogram of the FWHM values calculated using the two-point method on ATATS data (left) and PiGSS data (right).}
\label{fig:FWHM_hist}
\end{center}
\end{figure*}

Fig. \ref{fig:FWHM_hist} shows histograms of FWHM values calculated using the two-point method on the ATATS and PiGSS data.  We truncate the horizontal axis at a FWHM of $12^{\circ}$ for ATATS and $4^{\circ}$ for PiGSS, but the tail of large FWHM values continues as a result of the breakdown of Equation \ref{eqn:r_FWHM} as described in $\S$\ref{sec:two_point}.  While the distributions of FWHM values pictured in Fig. \ref{fig:FWHM_hist} are strongly peaked, they are not symmetric.  We therefore estimate the FWHM by taking the median of our results, and finding the extremes that encompass 68.3\% of our data.  We find that the mean FWHMs are $2.29 \pm 0.01^\circ$ for the ATATS data and $1.10 \pm 0.01^\circ$ degrees for the PiGSS data.

\subsection{$\chi^2$ minimization}
\label{sec:results:chi2}

\begin{figure*}[hbt!]
\begin{center}
\plottwo{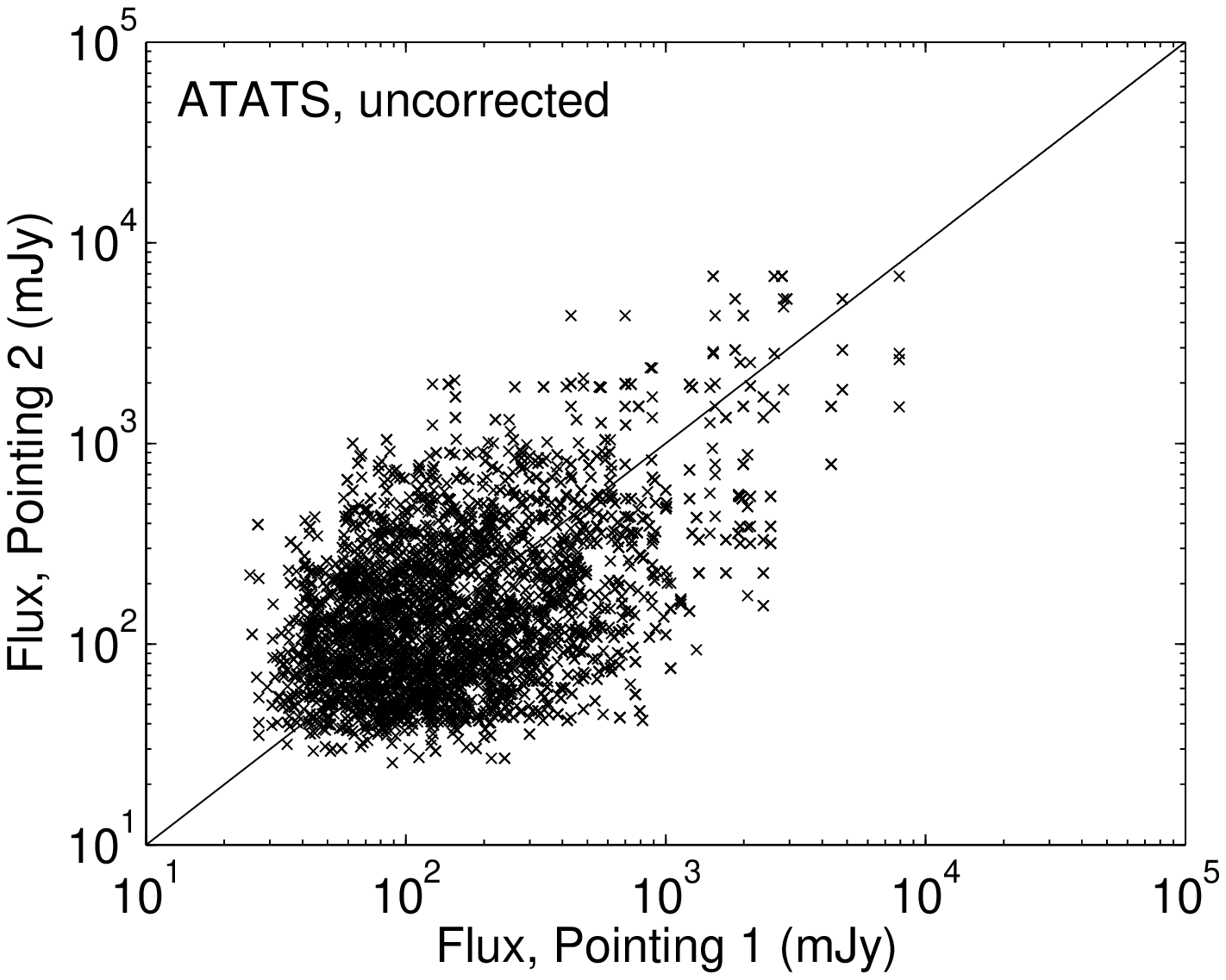}{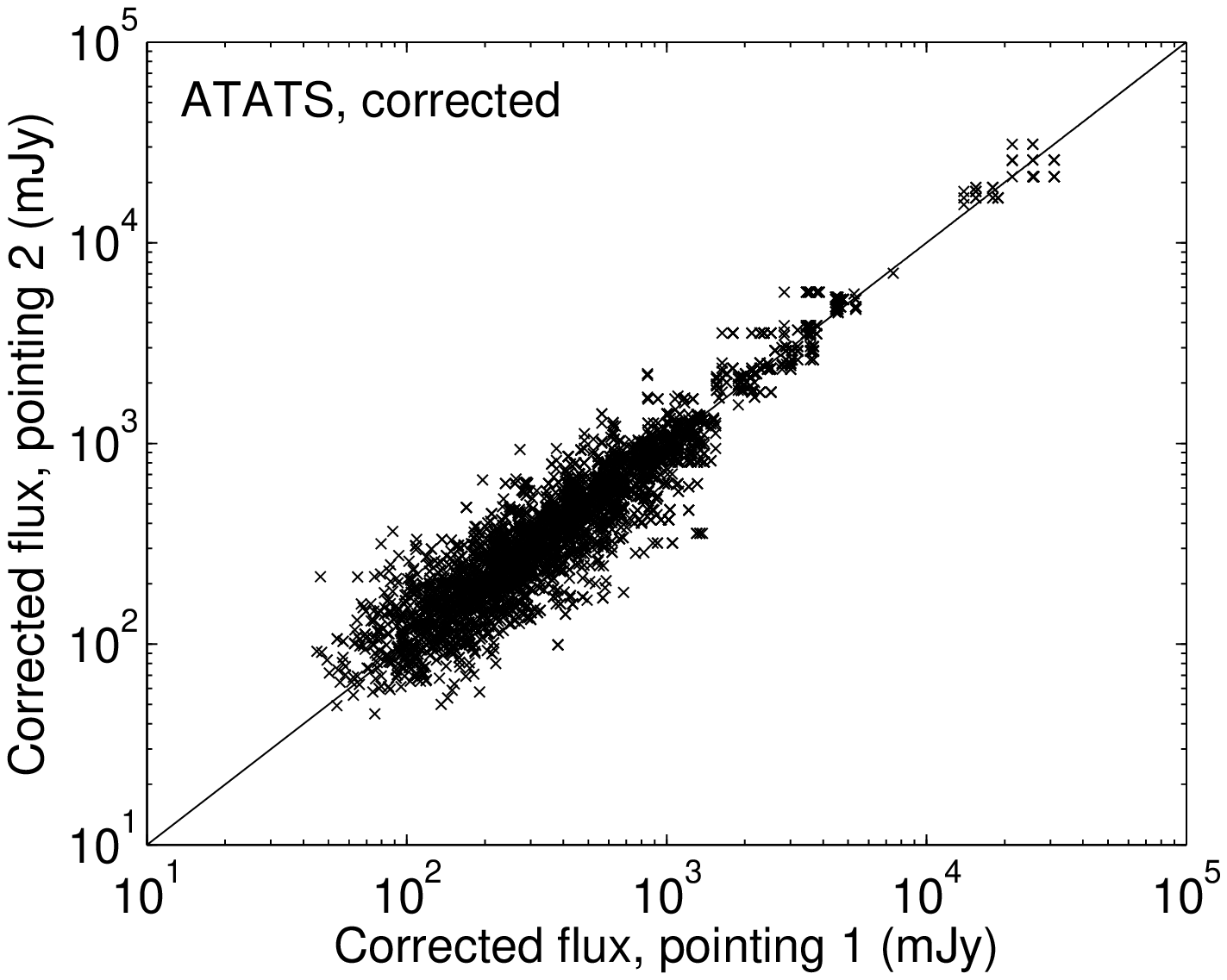}
\caption{$S_1$ vs. $S_2$ for identical ATATS sources, prior to (left) and after (right) correcting for the Gaussian beam attenuation using the best-fit (minimum $\chi^2$) value for the FWHM.  The appearance of clusters of points with the same flux density in either the vertical or horizontal direction results from the fact that some sources appeared in more than two pointings: to accumulate the maximum possible number of two-point matches, we match each appearance of a source with every other appearance.}
\label{fig:s1s2_ATATS}
\end{center}
\end{figure*}

\begin{figure*}[hbt!]
\begin{center}
\plottwo{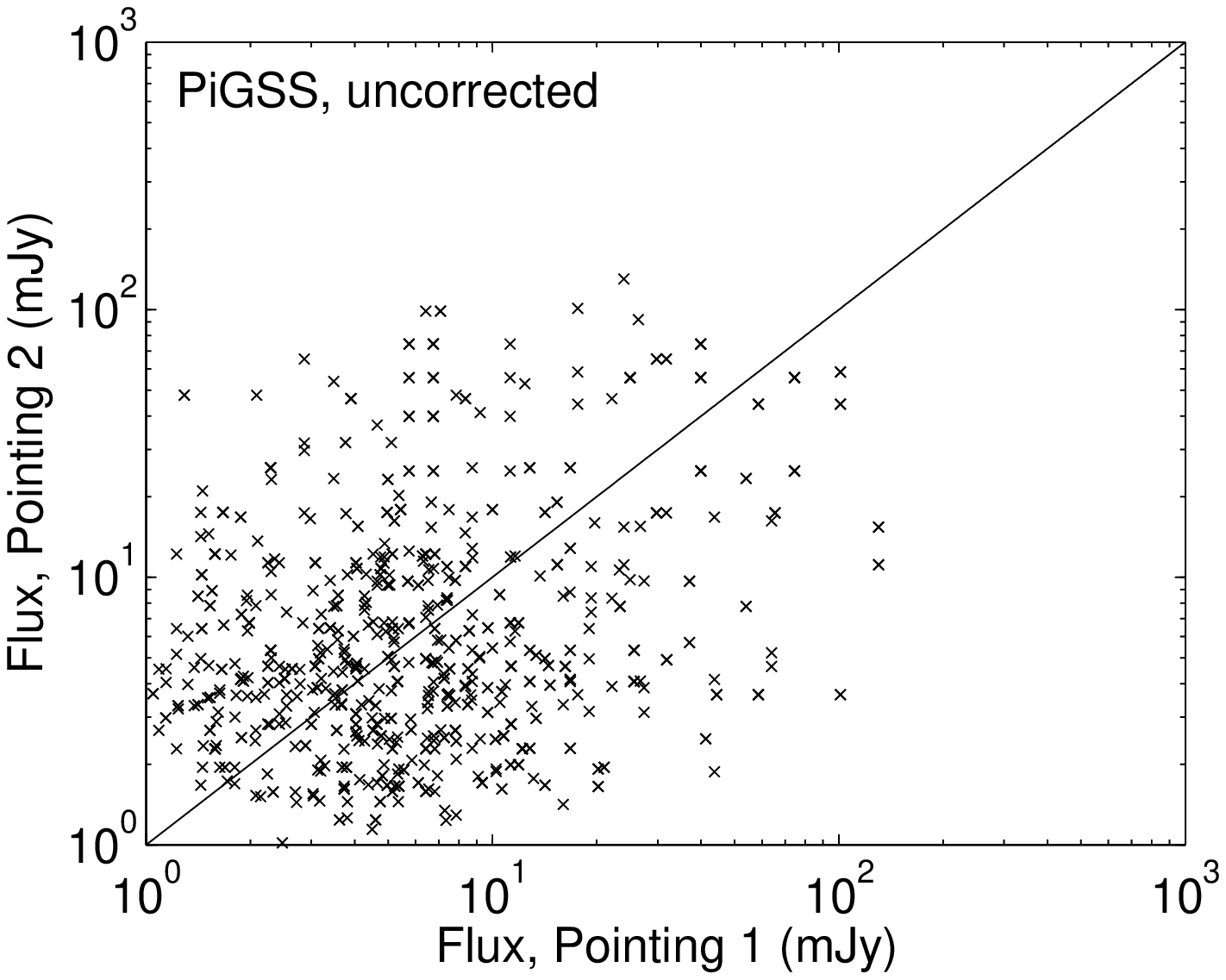}{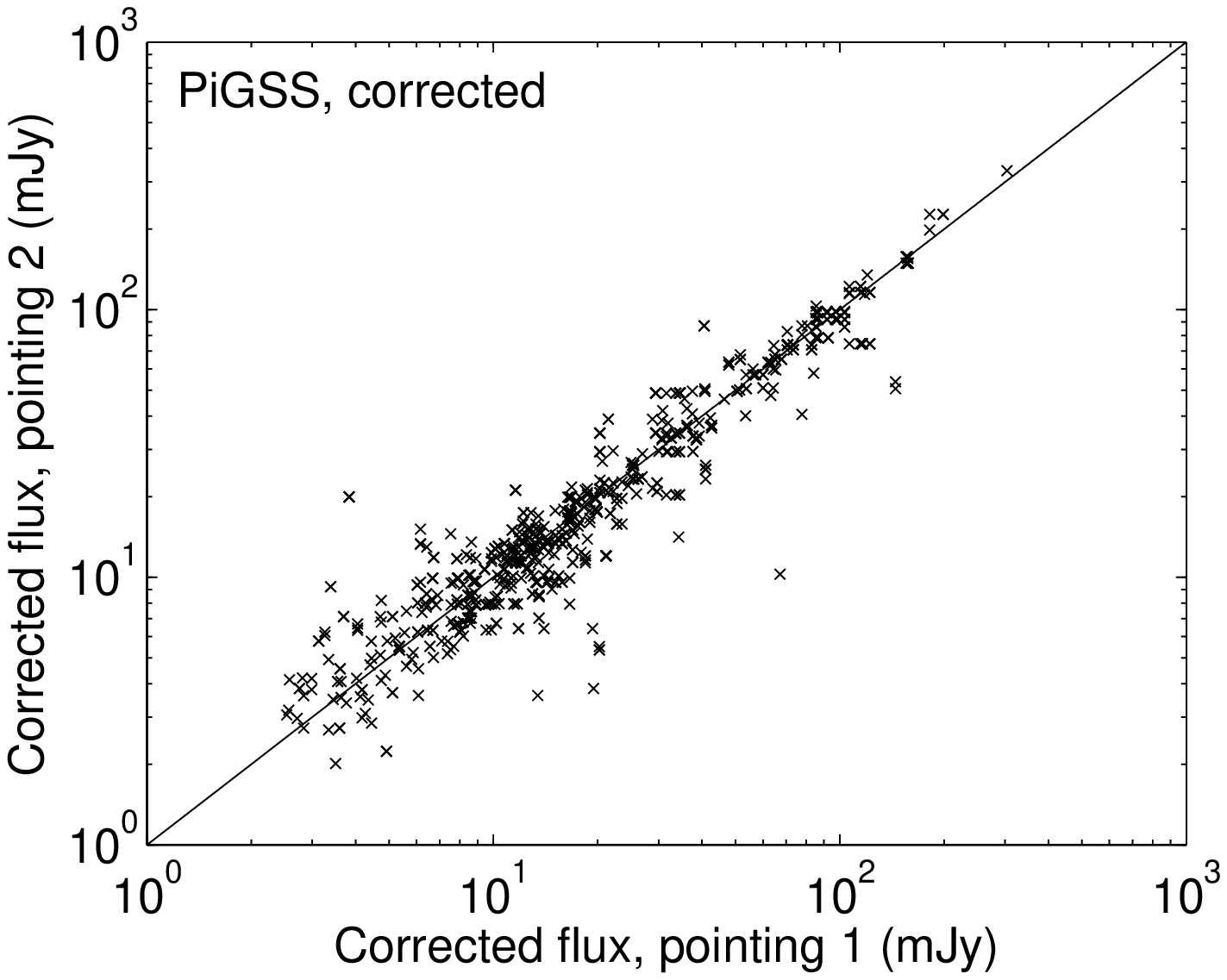}
\caption{$S_1$ vs. $S_2$ for identical PIGSS sources, prior to (left) and after (right) correcting for the Gaussian beam attenuation.}
\label{fig:s1s2_PiGSS}
\end{center}
\end{figure*}

If the assumption of a radial, Gaussian beam is correct, the values we calculate for $S_{1,\textrm{true}}$ and $S_{2,\textrm{true}}$ should be the same (and if our overall flux calibration is correct, should be equal to the true source flux density).  Fig. \ref{fig:s1s2_ATATS} shows $S_1$ vs. $S_2$ for both the uncorrected and corrected ATATS data.  Fig. \ref{fig:s1s2_PiGSS} shows the equivalent plots for the PiGSS data.

\begin{figure*}[hbt!]
\begin{center}
\plottwo{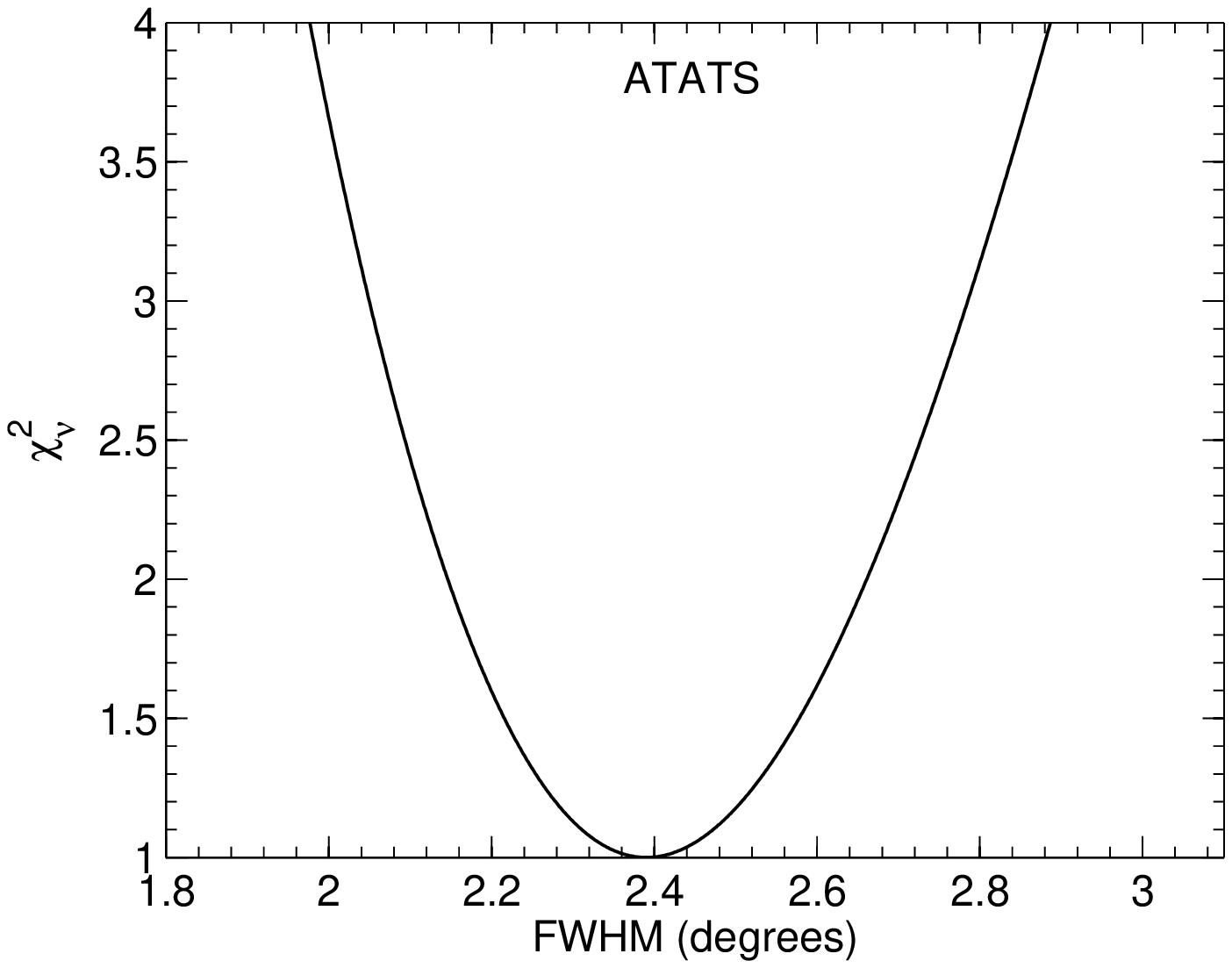}{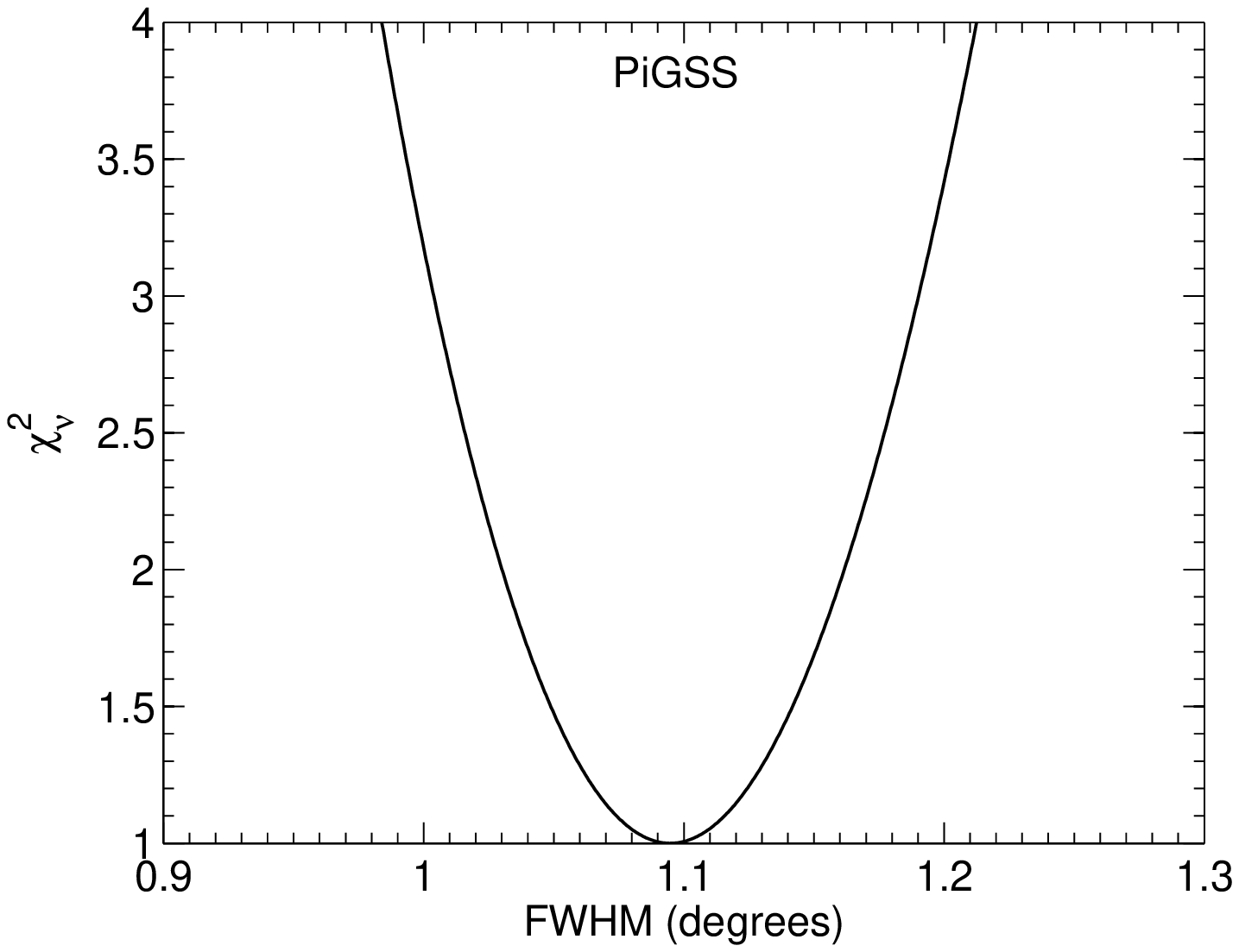}
\caption{$\chi^2_\nu$ vs. the FWHM value used to correct the ATATS and PiGSS data.  The uncertainties have been scaled up so that the curves both achieve a minimum of one.}
\label{fig:chi2}
\end{center}
\end{figure*}

Fig. \ref{fig:chi2} shows a plot of $\chi^2$ as a function of FWHM for both ATATS and PiGSS.  We had 4,303 and 823 degrees of freedom in our one-parameter fits for ATATS and PiGSS, respectively.  The minimal $\chi^2_\nu$ values for the fits are quite large: for the ATATS data the value is $\sim$21, and for PiGSS the value is $\sim$10.  This is not due to the differences between the two models from \citet{Welch:ATA_memo_66}, expressed in Equations \ref{eqn:bessel_beam} and \ref{eqn:gauss_beam}: after performing a $\chi^2$ analysis that corrected the flux densities using the beam shape from Equation \ref{eqn:bessel_beam}, we calculated similar values for $\chi^2_\nu$ for both ATATS and PiGSS.  

The large values of $\chi^2_\nu$ are due to a combination of systematic underestimation of uncertainties in flux density, the fact that the beam is not a perfect circular Gaussian \citep{Harp:ATA_beam_holography}, and the fact that there are probably a few mismatched sources.  This value could be reduced by fitting a more complex beam shape that includes beam ellipticity and beam angle, which neither Equation \ref{eqn:bessel_beam} nor Equation \ref{eqn:gauss_beam} took into account.  

\begin{figure*}[hbt!]
\begin{center}
\plotone{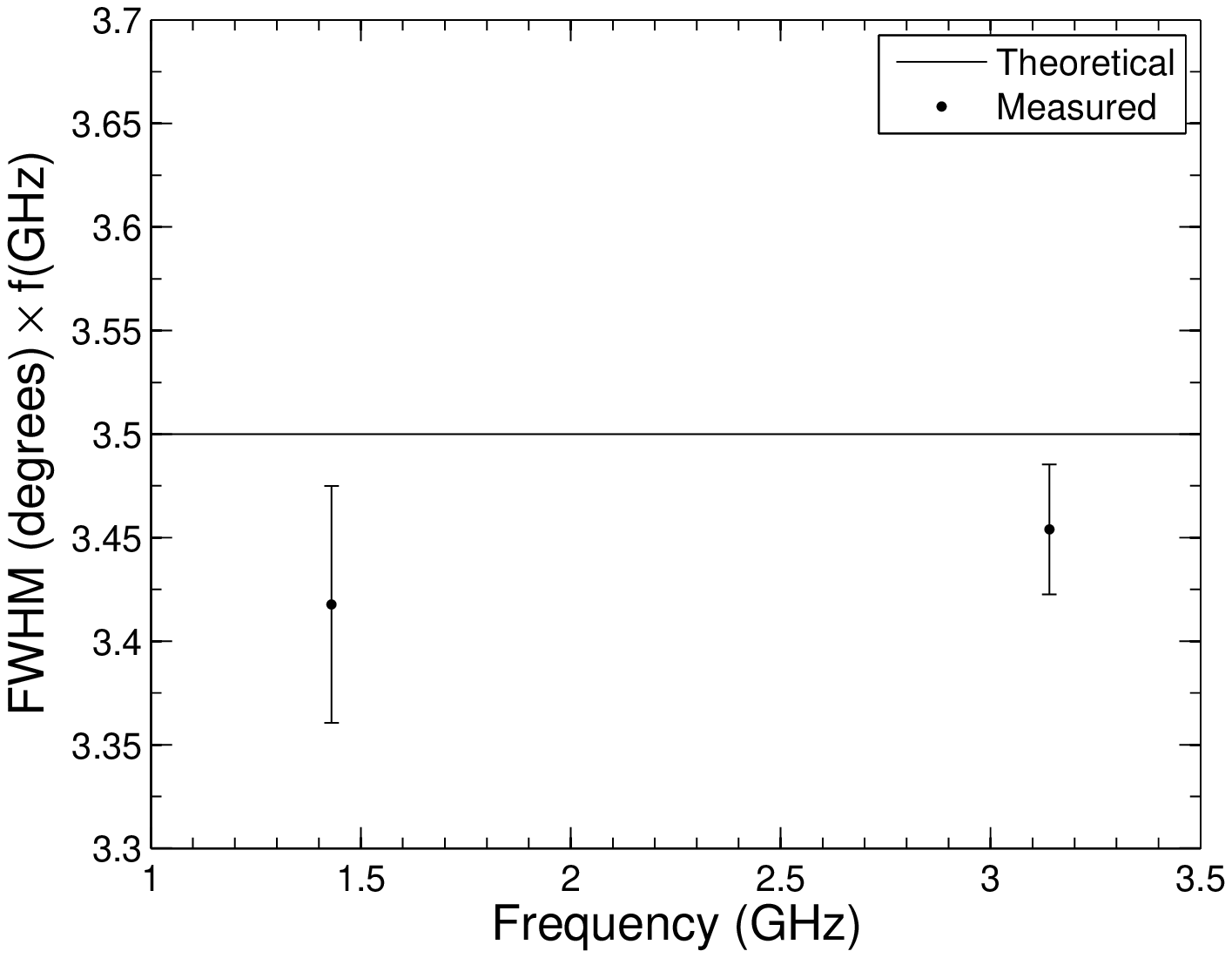}
\caption{The primary beam's FWHM $\Theta_{1/2}$ vs. observing frequency, shown over the relevant portion of the ATA's 0.5--10 GHz frequency range.  Points on the solid line are the expected values from Equation \ref{eqn:FWHM}, and the two large points are the FWHM values that we calculated using the $\chi^2$ method, as reported in $\S$\ref{sec:results:chi2}.}
\label{fig:FWHM_theo_meas}
\end{center}
\end{figure*}

After correcting the data and performing a $\chi^2$ minimization as described in $\S$\ref{sec:chi2}, we arrive at a best-fit FWHM value of $2.39 \pm 0.04^{\circ}$ for the ATATS data, and $1.10 \pm 0.01^{\circ}$ for the PiGSS data.  Fig. \ref{fig:FWHM_theo_meas} compares our results with the expression for expected FWHM, given in Equation \ref{eqn:FWHM}.

Clearly we are seeing a slightly narrower beam width than the width predicted by \ref{eqn:FWHM}.  The bias of a few percent is probably due to a combination of systematic errors including imperfect understanding of the ATA antennas, the inadequacy of the Gaussian beam assumption, and other losses that have a radial effect.

\section{Simulation: ATA-350 and SKA}
\label{sec:ska}

Here we use a simulation to investigate the dependence of the accuracy of the telescope's FWHM on antenna number.  As the number of antennas in an array increases, both the sensitivity of the array and the number of detectable sources increase.  Both of these factors allow us to measure the FWHM of the telescope's mean primary beam more accurately as the number of dishes increases. 

We assume each simulated array has a System Equivalent Flux Density, or SEFD, of $\sim$6000 (the SEFD of the ATA-42; see Equation \ref{eqn:radiometer}), a bandwidth $\Delta \nu$ of 200 MHz, and consists of identical 6.1 m antennas.  We generate simulated sources by assuming snapshot observations with an integration time $\tau$ of 1 minute, an observing frequency of 3.14 GHz (the same as the PiGSS data), and a power-law distribution of source density as a function of flux density \citep{Galactic_radio_astronomy}:

\begin{equation}
\frac{dN}{dS} \propto S^{-2} \,\, .
\label{eqn:dN_dS}
\end{equation}

Assuming the power-law distribution of source density from Equation \ref{eqn:dN_dS}, we arrive at the number of sources per steradian theoretically observable by a telescope sensitive to a given RMS flux density: 

\begin{equation}
\frac{N}{\Delta \Omega} = \frac{N_0 S_0^2}{\Delta S} \,\, ,
\label{eqn:source_density}
\end{equation}

\noindent
where $N$ is the number of detectable sources, $\Delta \Omega$ is the solid angle subtended by the telescope's field of view, $N_0 = 3 \times 10^6$ Jy$^{-1}$ ster$^{-1}$, $S_0 = 0.01$ Jy ($N_0$ and $S_0$ are both derived from the plot of source count vs. flux density at 1.4 GHz on p. 651 of \citet{Galactic_radio_astronomy}), and $\Delta S$ is the RMS flux density of the observation, which is dictated by the radiometer equation:

\begin{equation}
\Delta S = \frac{2 k T_{sys}}{A_{eff}\sqrt{N_A\left(N_A - 1\right) \tau \Delta \nu}} \,\, ,
\label{eqn:radiometer}
\end{equation}

\noindent
where $T_{sys}$ is the system temperature, $N_A$ is the number of antennas, $\tau$ is the integration time, and $\Delta \nu$ is the bandwidth.  The term $(2kT_{sys} \, / \, A_{eff})$ is also known as the SEFD.  (Note that the actual source count distribution as a function of flux is more complicated, and is only proportional to $S^{-2}$ over the $\sim$1--1000 mJy range of fluxes detected in the PiGSS survey.)

The RMS noise should decrease with an increasing number of antennas.  This fact, combined with the increasing number sources that will be detectable as RMS decreases, should result in better estimations of the beam's FWHM using the $\chi^2$ fitting method we describe in \ref{sec:chi2}.  To test our prediction, we simulate datasets from telescopes with an increasing number of dishes, to which we then apply our $\chi^2$ fitting method in the same way that we fit the real ATATS and PiGSS data.  We assume that the arrays in our simulations are not confusion limited.

We simulate observations from seven different arrays with numbers of antennas ranging from 42 to 2688 in powers of two.  For each of the seven arrays, we generate and analyze 1,000 different datasets.

For each simulated dataset, we generate a field of the appropriate number of sources by distributing them randomly throughout a 12.6 deg$^2$ field identical to the seven-pointing PiGSS field shown in Fig. \ref{fig:pointings}.  We then add Gaussian noise with a mean equal to the RMS of each source's flux density.  After distributing the sources across the field, we "observe" all of the sources in each of the seven pointings, where the observed flux density of the source is equal to the intrinsic flux density scaled down by the gain as a function of distance from the pointing center.  As we have throughout this work, we assume the primary beam is a circular Gaussian with a FWHM equal to 1.10$^\circ$, which we calculated in $\S$\ref{sec:results:chi2} after analyzing the PiGSS data.  We eliminate any observed sources whose fluxes have been scaled down to less than the RMS, and we only count 5$\sigma$ sources as detections.  After accumulating the lists of sources in each of the seven pointings, we use the same procedure that we use in $\S$\ref{sec:chi2} to determine the best-fit FWHM of the primary beam and its uncertainty.

\begin{figure}
\begin{center}
\plotone{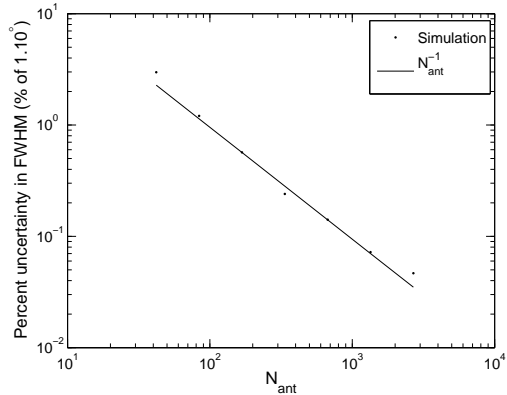}
\caption{The percentage uncertainty in the FWHM of the telescope's primary beam as a function of antenna number.  The solid line is a power law fit to the simulation results, and has an index of 1.}
\label{fig:simulation}
\end{center}
\end{figure}

We find that as the number of dishes increases, the uncertainty in the primary beam's FWHM decreases like a power law with an index of $\sim$1 (see Fig. \ref{fig:simulation}).  The first data point, corresponding the ATA-42, reports a FWHM uncertainty of 0.03$^{\circ}$, which is consistent to within a factor of three with the results from the PiGSS data reported in $\S$\ref{sec:results:chi2}.  As expected, our results return a median FWHM equal to 1.10$^\circ$, which is equal to the value we used when we simulated the data; and each simulation returns a $\chi^2_\nu$ value near 1, which is reasonable since we haven't introduced any systematic errors into our calculations.

The decreasing error in the calculated FWHM of the beam with increasing dish number will be important for future telescopes with large numbers of dishes.  For example, with the SKA-3000 \citep{SKA_memo_100}, one will be able to characterize the primary beam much better by using images taken with a fraction of the integration time characteristic of the above ATA-42 observations.  Extrapolating the power law in Figure \ref{fig:simulation} to an array with 3000 antennas, we can see that in the absence of systematic errors one could estimate the FWHM of the SKA-3000 to within 0.02\%.

\section{Conclusions}

While consistent with our $\chi^2$ minimization results, the method of calculating the beam's FWHM using the two-point method has systematic problems, including the fact that the best flux density-distance pairs lead to the least reliable FWHM values.  We also find the $\chi^2$ minimization to be superior to the two-point method because of its lack of systematic problems, its ability to use all of our data in the fit, and its compatibility with more complex models.

Overall, both our analysis of real ATA data and our simulation confirm that the theoretical FWHM values from \citet{Welch:ATA_memo_66} are reasonable, as they are consistent with our results at the frequencies of both the ATATS and PiGSS surveys.  The theoretical values are also consistent with the results from radio holography \citep{Harp:ATA_beam_holography} and off-axis measurement of point sources \citep{Macmahon:ATA_memo_83}.

The results of our simulation, which calculates the uncertainty in the FWHM of the primary beams of telescopes with different numbers of dishes, will be important for the primary-beam characterization of future LNSD telescopes such as the ATA-350 and the SKA.  The dramatically increased number of sources detectable by these telescopes will allow much better constraint of beam properties, thus improving the accuracy and fidelity of mosaicked images.  \citet{1993A&A...271..697C} note that in the absence of systematic errors, one can achieve a dynamic range in a mosaicked image of $\sim$3,000 and a fidelity index of $\sim$150 by knowing the primary beam to an RMS error of $\sim$1\%.  We can measure the beam shape of the ATA to within a tenth of a percent, suggesting that we have the potential to achieve extremely high dynamic ranges and fidelity indices in ATA mosaics.

An extension of our model could be developed to allow fitting not only of the beam's FWHM, but also of its ellipticity and its angle on the sky.  Additional effects such as spillover from the ground, mechanical deformation of the dish, frequency and baseline dependence of the beam, and time variability of sources are surely factors to be considered; however, most of these are probably relatively minor effects when compared with the effect of using a simple, Gaussian model for a beam that we know is non-Gaussian at some level.  Regardless, all of these effects should be taken into consideration in future primary beam studies on telescopes such as the ATA-350 the SKA, and other radio interferometers.

The authors would like to acknowledge the generous support of the Paul G. Allen Family Foundation, who have provided major support for design, construction, and operations of the ATA. Contributions from Nathan Myhrvold, Xilinx Corporation, Sun Microsystems, and other private donors have been instrumental in supporting the ATA. The ATA has been supported by contributions from the US Naval Observatory in addition to National Science Foundation grants AST-050690, AST-0838268, and AST-0909245.

We would also like to acknowledge the advice and guidance of Dick Plambeck, Mel Wright, Mike McCourt, Jonathan Pober, Adam Morgan, James McBride, and the members of the Berkeley Radio Astronomy Lab in the preparation of this paper.  CLHH is supported by a UC Berkeley Chancellor's Fellowship and an NSF Graduate Fellowship.

\bibliographystyle{apj}
\bibliography{/Users/Chat/1_Berkeley/Research/LaTeX/chat_refs}

\end{document}